# Deep Learning-Enabled Dissolved Oxygen Sensing in Biofouling Environments for Ocean Monitoring

**Authors:** Nikolaos Salaris[1,2], Adrien Desjardins[2,3], Manish K. Tiwari[1,2*]

[1] Nanoengineered Systems Laboratory, Mechanical Engineering Department, University College London, London WC1E 7JE, UK

[2] UCL Hawkes Institute, University College London, London W1W 7TS, UK

[3] Department of Electrical and Computer Engineering, University of British Colombia, 5500-2332 Main Mall, Vancouver, BC Canada V6T 1Z4

*Corresponding author details

Email: m.tiwari@ucl.ac.uk, Phone: +44 20 3108 1056 (Manish K. Tiwari)

## Abstract

The escalating climate crisis and ecosystem degradation demand intelligent, low-cost sensors capable of robust, long-term monitoring in real-world environments. Absolute dissolved oxygen (DO) concentration is a key parameter for predicting climate tipping points. Inexpensive optoelectronic sensors based on microstructured polymer films doped with phosphorescent dyes could be readily deployable; however, signal drift and marine biofouling remain major challenges. Here, we introduce a sensing paradigm that combines camera-based DO sensors with a visual transformer (ViT)-based physics-informed neural network (PINN) for high-fidelity sensing under biofouling conditions. Training and testing data were obtained from an algae-laden water tank over 14 days to capture accelerated biofouling. The ViT-PINN, which embeds the Stern-Volmer (SV) equation into the loss function, reduces mean average error (MAE) by 92% and 89% compared to classical statistical and ML approaches, achieving ~2 µmol/L absolute error. A deep ensemble further quantifies predictive uncertainty, enabling self-diagnostic sensing.

## Main

The intensifying pace of climate change and anthropogenic pressure on aquatic ecosystems has created an urgent, global need to widen environmental monitoring [1]. Oceans in particular are severely under-measured. To this end, Dissolved Oxygen (DO) is an Essential Ocean Variable by global observing systems [2,3]. Specifically, steady warming of oceans is driving a widespread decrease in DO, a phenomenon known as ocean deoxygenation, that severely disrupts marine ecosystems and can trigger the collapse of entire fisheries [4,5]. This makes DO a potent early warning indicator for the crossing of critical climate tipping points [6], with the expansion of hypoxic "dead zones" representing catastrophic ecological thresholds for marine life.

Our ability to forecast climate events is critically undermined by a lack of robust, widespread measurements. Current DO models can deviate from observed deoxygenation rates by as much as 100% - a profound uncertainty that can only be reduced by a new generation of sensors capable of providing high-resolution, long-term measurements in the field [4]. Simultaneously, monitoring algal biomass is vital for tracking harmful algal blooms [7] and for optimizing industrial applications like carbon capture and biofuel production that depend on robust, affordable sensing [8]. However, deploying sensors for long-term, autonomous in-situ measurements is severely hampered by a universal challenge: biofouling [9]. The inevitable colonisation of the sensor surfaces by microorganisms alters their response, leading to signal drift and catastrophic failure, necessitating costly and frequent maintenance. Specifically, these issues hamper optical and electrochemical probes [10,11].

To combat biofouling, significant effort has been invested in materials-based solutions designed to prevent organism attachment, such as hydrophilic and hydrogel-based coatings [12], bio-inspired and biomimetic coatings [13], copper-free and metal-free hybrid coatings [14], biocide-leaching coatings [15,16], and non-toxic, low-surface-energy polymers [17]. Nonetheless, the eventual development of a resilient microbial fouling layer limits the efficacy of all such methods [18,19]. Critically, for luminescence-based methods, the application of an anti-fouling layer presents a fundamental trade-off with sensor performance. One such example is optical DO sensing via quenching of phosphorescence; a promising alternative to traditional electrochemical probes, offering high sensitivity, avoiding oxygen consumption and large drifts [20]. These sensors rely on the Stern–Volmer (SV) principle, where the luminescence of an embedded indicator is dynamically quenched by oxygen. For such sensors, applying anti-biofouling coatings can introduce optical interference and corrupt the sensor kinetics [21]. Moreover, the real-world application of such DO sensors is constrained by expensive instrumentation. Therefore, the inherent trade-off associated with anti-biofouling coatings and the high cost of current solutions necessitate a shift towards intelligent sensing. This would enable a practical deployment of self-contained sensor packages at scale for long periods of time and high geographic coverage.

Traditionally, phosphorescence-based DO measurements have been quantified by a single, spatially-averaged measurement from a photodetector, but recently the use of cameras has provided 2D analyte maps that allow for a better understanding of the underlying physical processes [20, 22–30]. In addition, the SV relationship, which underpins luminescence-based oxygen quantification, carries the implicit assumption that the sensing matrix is spatially homogeneous [23]. While this a reasonable approximation in controlled laboratory conditions, it collapses in real-world deployments, where spatially heterogeneous processes such as biofouling, patchy film degradation, and uneven illumination are unavoidable [9, 24, 28]. Crucially, in the case of the algal growth the heterogeneity is not static: the spatial distribution of fouling evolves continuously, meaning that the sensor regions yielding the most reliable signal at one time point may be among the most degraded at another. This non-stationarity renders any calibration strategy based on a fixed selection of pixels fundamentally brittle over extended deployments.

This fundamental mismatch between the mathematical framework of calibration and the heterogeneity of luminescence-based sensors poses an obstacle to robust, long-term real-world applications. Realizing the full potential of imaging-based sensors therefore demands a shift, steering away from global calibration curves and going towards intelligent frameworks that resolve the governing physics locally, distinguishing valid signal from artifact on a pixel-by-pixel basis. To this end, Physics-Informed Neural Networks (PINNs) can assimilate physical laws as a powerful inductive bias, promoting parsimony and improving generalization, especially with noisy, real-world data [31]. Fundamentally, this acts like a principled regularizer: it lowers data requirements, stabilizes training in the presence of noise, and reduces overfitting to spurious correlations when compared to purely data-driven models [32,33].

Here, we address the challenge of optical DO sensing via quenching of phosphorescence in a realistic (noisy) environment by introducing a low-cost intelligent sensing framework that is robust to spatio-temporal heterogeneities. Our approach integrates three key innovations: 1) a novel, highly scalable oxygen-sensitive film based on a PtOEP–polystyrene composite, fabricated using a simple, self-assembly dip-coating method [34]; 2) a low-cost, automated imaging system using a Raspberry Pi and a UV LED; and 3) a computational framework built on PINNs and spatial feature extractors. This framework moves beyond simple calibration by directly embedding the SV equation into the training process and applying ML techniques towards an imaging-based optical sensor. We validate our approach against a comprehensive hierarchy of methods, starting from classical pixel-averaging and "best pixel" selection to sophisticated, feature-engineered gradient boosting and ViT-based models. We demonstrate through direct comparison that our ViT-based PINN architecture is more accurate, with a 92% lower MAE of 2 µmol/L (when trained in chronological order) across 14 days of accelerated biofouling, compared to the typical average-based calibration. This methodology allows high spatio-temporal resolution of the sensor in biofouling conditions. Crucially, while using relatively simple hardware and components, the robustness of our framework implies that it could serve as a blueprint for resilient, self-diagnosing, and scalable optical sensors, which are crucial for monitoring climate tipping points.

## Results

**Limitation of typical Averaging and Physics-Based Methods**

To evaluate our sensing frameworks under realistic fouling conditions, we constructed a low-cost imaging platform comprising a Raspberry Pi camera, a 395 nm UV LED, and a PtOEP–polystyrene oxygen-sensitive film fabricated by dip-coating (see Methods). The sensor was submerged in a fertilised algae-laden water tank and calibrated daily over a 14-day period against a commercial Pyroscience DO probe, providing ground truth across five oxygen concentration ranges **(Fig. 1)**. This accelerated biofouling protocol allowed us to systematically assess model performance as the algae colonised both the sensor surface and the surrounding water column, progressively degrading the optical signal through direct surface occlusion.

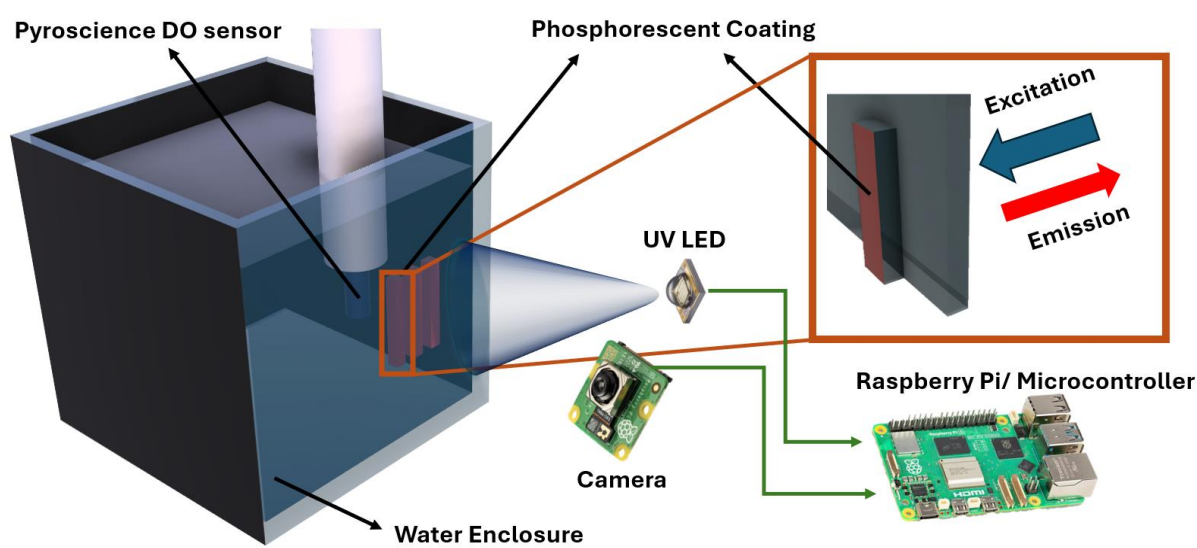

**Figure 1:** Experimental Setup

Schematic depiction of the experimental setup including: a UV LED, a water enclosure, the industry DO sensor, the phosphorescent film, the Raspberry Pi microcontroller. An emphasis is given on the placement of the film where the direction of the film was on the other side of the UV transparent Acrylic that was submerged in the water tank.

The first approach was to apply a 'Global Average' model (GA), which simulates a standard single-point calibration by fitting the Stern-Volmer (SV) equation to the mean red intensity of all pixels (for each frame) [30]. This is equivalent to the single-point measurements used as an industry standard when calibrating the sensor. This proved highly inaccurate for predicting dissolved oxygen concentrations in a biofouling environment, resulting in a high cross-validated Mean Absolute Error (MAE) of 28.2 µmol/L. Critically, when training and testing all models, cross-validation was performed by holding out an entire experimental day as the test set in each fold (Leave-One-Out Cross-Validation) rather than randomly partitioning individual observations across days (see Methods). This choice guards against the temporal autocorrelation and covariate leakage that random subsampling would introduce, since measurements acquired on the same day share environmental and biofouling conditions. This provides a stringent test of the model's ability to generalise to genuinely unseen temporal conditions.

We then explored a more sophisticated 'Best Pixels' strategy. In this approach, each of the 2,304 pixels (each image was down sampled to 48x48 pixels from 1920x1080) was individually fitted to a linear and non-linear (two-site approach) SV equation across the training data to derive the fitted parameters and a suite of performance metrics from physics-derived statistics ("Per Pixel Fitting" strategy). Specifically, we systematically selected the top 10, 100, 1,000 and all 2,304 performing pixels based on ranking criteria derived from the calibration procedure and based on their physics-derived parameters, including: the quenching constants ($K_{SV}$), the Dynamic Range (*DR*), the zero-oxygen intensity ($I_0$), the Limit of Detection (LOD) and the goodness-of-fit ($R^2$) to a linear and non-linear (two-site) SV equation (see equations 4 to 8 for definitions in the Methods section). The red-channel intensities of these best performing pixel cohorts were averaged to create a robust 'super-pixel' signal for prediction (one cohort for each metric). This naive feature selection offered a marginal improvement, with the best-performing strategy ('Top 10 by $R^2$') (R210NL) achieving an MAE of 23.4 µmol/L. Both linear (1-site) and non-linear (2-site) SV models were evaluated for all strategies, with the non-linear implementation providing slightly better but still insufficient results; MAE of 23.4 µmol/L versus 26.0 µmol/L, for the best non-linear and linear strategy, respectively. Nonetheless they fundamentally remained inaccurate.

Additionally, we tested a physics-reinforced machine learning approach using a LightGBM (LGBM) model [35], which was selected for its computational efficiency when handling large datasets. Here the physical quantities are supplied as input features to the model without enforcing physical consistency during training, which is distinct from a PINN where the governing equation is embedded directly in the loss function. This model was supplied with a rich input feature set that included the red channel intensity statistics (average for each frame and standard deviation for each oxygen concentration measurement interval) but also their aggregated physical parameters (mean $I_0$, mean $K_{SV}$, mean *LOD*, mean *DR*, mean $R^2$). This hybrid approach (LGA) yielded a further improvement, reducing the MAE to 18.5 µmol/L.

Nonetheless, the poor performance for both approaches was quantitatively confirmed with parity plots where predictions from both the typical GA and the physics-reinforced LGA models show significant deviation and wide scatter relative to the line of perfect agreement **(Fig. 2A-B)**. Furthermore, the residual plot, which shows prediction error versus the ground truth DO, reveals that the error magnitude is not constant, indicating that the model struggles to perform consistently across the operational range of DO concentrations tested with wide fluctuations **(Fig. 2C)**. Additional analysis of the LGA method highlights once more its unreliability. The error distribution histogram is broad, with a large standard deviation (σ), indicating a wide range of prediction errors rather than a consistent, predictable offset **(Fig. 2D)**. These results collectively demonstrate that methods reliant on spatial averaging or static feature selection are fundamentally brittle with high prediction MAE **(Fig. 2E, F)**. In essence, they fail to account for the dynamic, non-uniform nature of biofouling, where the 'best' performing sensor regions are not static. This establishes the critical need for an architecture that can explicitly learn and adapt to the spatiotemporal heterogeneities of algal growth on the sensor surface.

The underlying cause for the failure of these averaging-based methods is revealed by analysing the spatial distribution of the sensor's response. Heatmaps of the pixel-wise goodness-of-fit ($R^2$) to the SV equation show that there is no strong physical correlation between the predictors and the underlying film phosphorescence. Specifically, non-uniformly distributed regions of the parameters were found both within and outside the boundaries of the sensor film for linear and non-linear models **(Fig. 3)**. This pronounced and erroneous spatial heterogeneity does not represent the underlying phosphorescent emission properties of the films and becomes apparent when compared to the raw RGB frame of the sensor. This means that these methods fail to represent the sensing film in the spatial domain using simple fitting and averaging techniques.

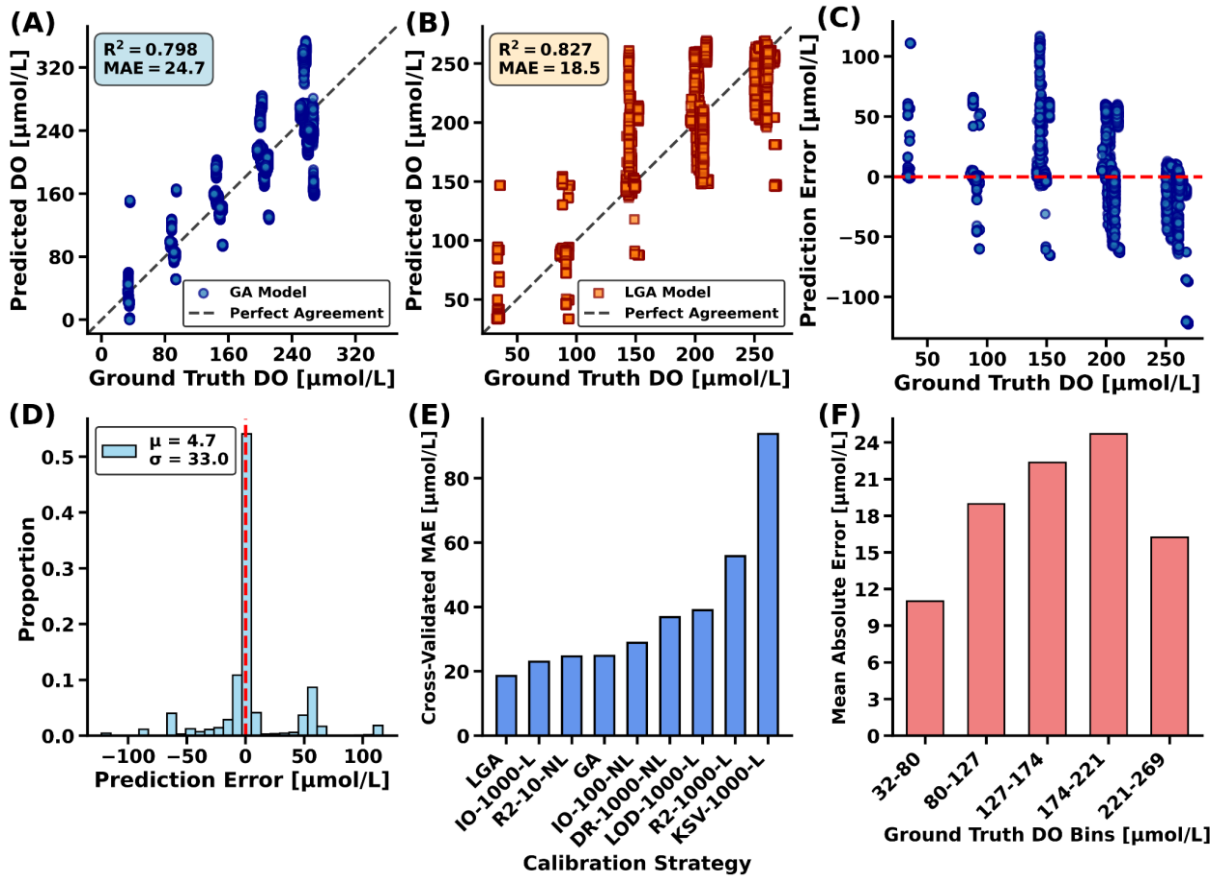

**Figure 2**: Failure of Conventional and Physics-Reinforced ML Methods in the Presence of Biofouling.
Parity plots of **(A)** the 'Global Average' (GA) model and **(B)** the best-performing LGBM model using aggregated physical parameters (LGA). The annotation box displays the $R^2$ and MAE (in units of μmol/L) for each model as metrics of prediction accuracy. Plots of **(C)** the Prediction errors versus the ground truth oxygen concentration and **(D)** the histogram of the distribution of Prediction Errors (a red line shows the point where the prediction error is 0). The y-axis shows the normalised frequency (Proportion), and the inset legend box provides the mean (μ) and standard deviation (σ) of the error. **(E)** Bar chart comparing the performance of conventional methods using the cross-validated test MAE as a metric. This included the average-based methods in tandem with the 'Best Pixels' strategies, including GA, LGA and the best-performing 'Best Pixels' strategies with different cohorts of top-performing pixels using specific physics-derived metrics. The "L" or "NL" in the name of the models corresponds to using a Linear and a Non-Linear SV fit, respectively. Specifically, these models are (left to right): i) **IO-1000-L** (top 1000 pixels selected by their zero-oxygen $I_0$ intensity); ii) **R2-10-NL** (top 10 pixels selected by their goodness-of-fit $R^2$); iii) **IO-100-NL** (top 1000 pixels selected by their zero-oxygen $I_0$ intensity ; iv) **DR-1000-NL** (top 1000 pixels selected by their Dynamic Range); iv) **LOD-1000-L** (top 1000 pixels selected by their Limit of Detection); v) **R2-1000-L** (top 1000 pixels selected by their $R^2$); and vi) **KSV-1000-L** (top 1000 pixels selected by their quenching constant $K_{SV}$). **(F)** A bar chart of MAE against the ground truth DO for the 5 main DO concentration ranges during the calibration procedure using the LGA model and their standard deviation in terms of MAE.

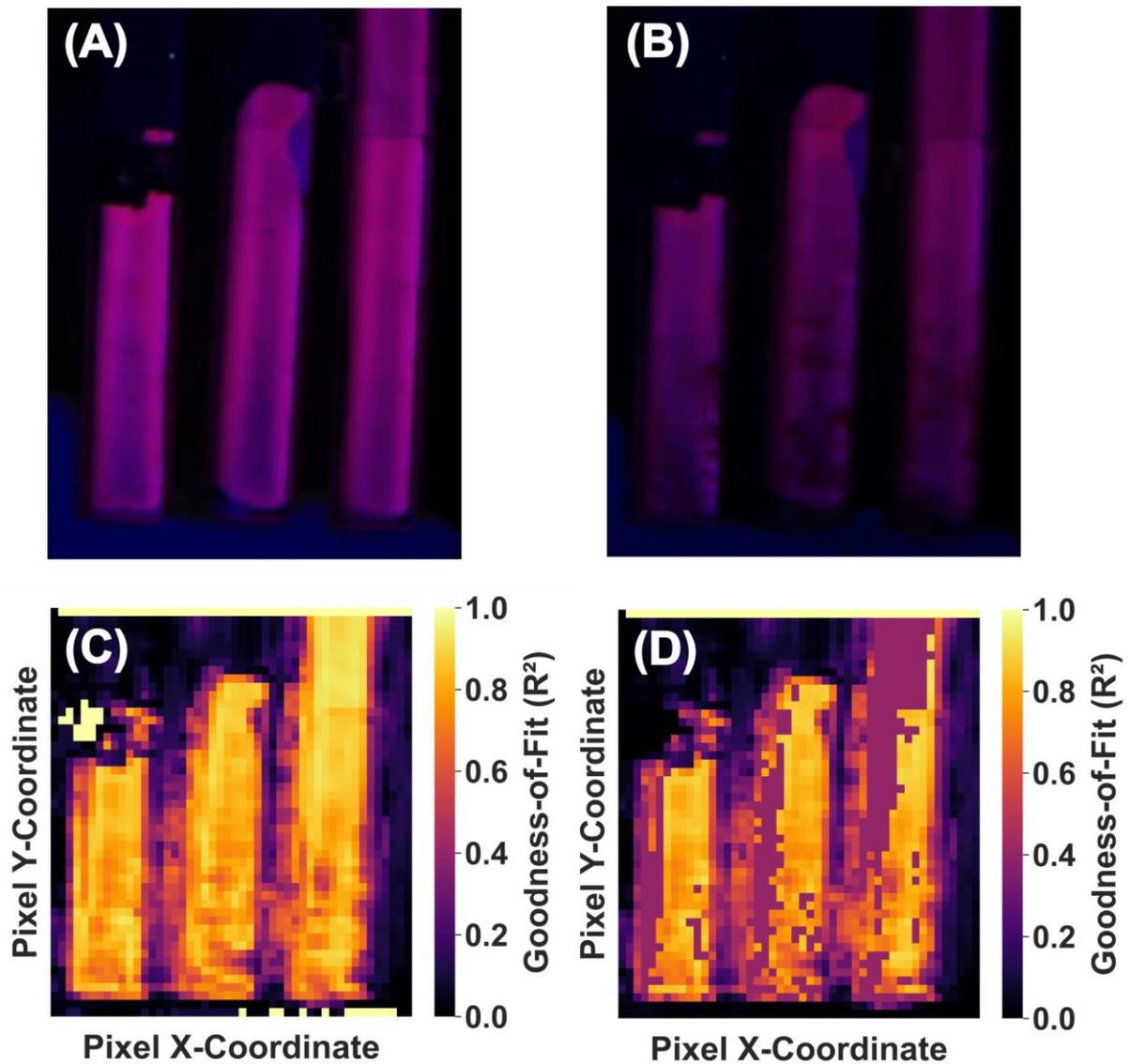

**Figure 3**: Spatial Heterogeneity and failure of "Best Pixels" approach.
Images of the unprocessed frames at two different fouling and oxygen levels: **(A)** 100 % V/V $N_2$ and 0% V/V% $O_2$ taken at the first day prior to algal growth and **(B)** 95% $N_2$ and 5% $O_2$ V/V at the 5$^{th}$ day, where there is significant surface occlusion due to algal growth. Heatmaps showing the per-pixel fitted parameter of the goodness-of-fit ($R^2$) to the linear **(C)** and non-linear **(D)** SV equations, averaged across all cross-validation folds to show the spatial heterogeneity of the sensor response.

**Spatial Information with Machine Learning: Improvement Without Understanding**
To establish a rigorous ML baseline, we evaluated a hierarchy of LightGBM (LGBM) and Convolutional Neural Network (CNN) models, with hyperparameters optimised using Optuna [36] (full details in Supplementary Section S1). We first tested position-agnostic LGBM models that treat each pixel as an independent sensor: a) a baseline using RGB values and temperature (LRGBT; MAE 41.8 µmol/L), b) a variant augmented with per-pixel non-linear SV parameters (LRGBTSV; 34.6 µmol/L), and c) a fully-featured version including all physics-derived statistics (LRGBSVTP). Incorporating pre-computed physical features yielded only marginal gains. Preserving pixel location through a position-aware LGBM, where the flattened 48×48×3 image and SV parameters served as input features (LSSV; 13.2 µmol/L), delivered a substantial improvement, though appending additional physics-derived statistics (LSSVP; 13.2 μmol/L) offered no further benefit. Downsampling the resolution had minimal impact (13.6 μmol/L at 64×64) but significantly increased computation time. The next step was to train CNN where we provided the full 48x48x3 RGB frames as input, together with temperature. Replacing the flattened representation with a CNN (ResNet-18 with CBAM attention), which directly processes the full 2D image grid to capture spatial correlations, yielded a further leap, reducing the MAE to 8.1 μmol/L (**Fig. S1A,** Supplementary Information Section S1).

In detail, the CNN's predictions clustered tightly along the line of perfect agreement (y=x), in stark contrast to the wider scatter of the best LGBM (LSSVP), particularly at low DO concentrations (**Fig. S1B**). Permutation importance analysis revealed that the position-aware LGBM relies on raw intensities at fixed pixel positions (**Fig. S1C**), making it inherently brittle to the non-stationary fouling patterns discussed above, while for the position-agnostic models' importance was distributed across temperature, RGB values and physics statistics (**Fig. S1D**). However, this broader feature reliance still proved ineffective without spatial relationships between pixels. This brittleness was further confirmed by heatmaps of the pixel-wise MAE for the position-agnostic models (LRGBT, LRGBTSV, LRGBSVTP), which revealed distinct regions of persistent high error across the sensor surface, demonstrating that adding physics-informed features may alter the spatial error pattern but cannot resolve the underlying heterogeneity (**Fig. S2**).

These results collectively establish two key findings: i) spatial awareness, defined as the ability to learn correlations between pixels rather than treating them independently, is the dominant factor in prediction accuracy; and ii) that the CNN, despite its superior performance, operates as a black box whose learned attention maps failed to align with the known spatial distribution of the phosphorescent film, learning implicit correlations without explicit knowledge of the quenching physics. This motivates the physics-informed approach that follows.

**Physics-Informed Neural Network Achieves State-of-the-Art Accuracy and Parsimony**
The next step was to constrain the deep convolutional neural network with a physics-based loss, so that the model "learns" to spatially identify regions where quenching of phosphorescence is occurring due to oxygen presence. By penalizing departures from the SV physics alongside standard supervision losses, the network learns to separate true quenching dynamics from artifacts. In detail, the core innovation of the PINN, as it is implemented here, is its dual-objective loss function. The network is not trained solely to match ground truth data but is simultaneously constrained to obey the governing physical laws of the system. The total loss, which the model seeks to minimize, is a weighted sum of a data-driven loss, and a physics-based loss (1). This approach was further enhanced by a third loss component, which provides direct supervision for the model's internal estimation of biofouling against labelled ground truth (2). In the latter, the total loss that the model seeks to minimize, is a weighted sum of a data-driven loss, a physics-based loss, and a biofouling supervision loss, representing a more constrained version of the base model. The biofouling supervised values came directly from the crystal violet measurements using UV-Vis spectrometry.

$$L_{total} = L_{data} + \lambda_p \cdot L_{physics} \quad (1)$$

and

$$L_{total} = L_{data} + \lambda_p \cdot L_{physics} + \lambda_b \cdot L_{biofouling} \quad (2)$$

Where (see Supplementary Information Sections S2, S3 for full details):

- $L_{data}$ is the supervision loss, which measures the discrepancy between the model's final prediction and the ground truth sensor reading.
- $L_{physics}$ is the physics residual, which quantifies how well the sensor's response adheres to the SV equation.
- $L_{biofouling}$ is the new supervision loss, which is the MSE between the predicted biofouling score and the normalised ground truth.
- $\lambda_p$ and $\lambda_b$ are hyperparameters that balance the relative importance of the data fidelity represented as $L_{data}$ and the physical consistency terms $L_{physics}$ and $L_{biofouling}$.

The physics loss is defined as the residual between the physics-predicted DO values ($[O2]_{predicted}$) and the ground truth ($[O2]_{ground\ truth}$) DO concentration (which is different for every pixel). For a linear version of the SV equation this would mean that the residual is defined as (3).

$$Lphysics = [O2]_{predicted} - [O2]_{ground\ truth} = \left(\frac{I_0}{I}\right) - 1 - K_{SV} \cdot [O_2]_{Ground\ Truth} \quad (3)$$

Importantly, the source of non-uniformity in the performance of the pixels corresponding to the phosphorescent film was associated with uneven: a) biofouling, b) illumination, c) application of the coating, and d) degradation of the phosphorescent material (such as photobleaching). In essence, the real-world conditions result in deviations from spatial homogeneity. In this direction, the resulting residual maps (error between the physics part and the ground truth data) double as self-diagnostic indicators, highlighting where the physical principles hold and where they deviate. This yields a system that is both more accurate and more interpretable, a fundamental advantage of applying PINNs. Specifically, by mapping the pixel-wise physics loss back to the image space, we can visualize the model's reasoning, which in turn informs us about the spatiotemporal properties of the sensor. The physics-based loss term is not only a regularizer; it is a powerful diagnostic tool. In effect, the PINN turns physical consistency into a more generalizable learning signal, enabling robust performance under drift and scarcity of labelled ground truth for spatially resolved measurements.

In particular, for any given input image, we can compute the pixel-wise SV equation residual (3). This residual map directly visualizes where and how much the sensor's response deviates from the expected physical model **(Fig. 4A)**. Essentially, it identifies regions of high error where the pixel intensities I (x,y) are not accurate oxygen sensors. In practice, the model has learned a "Physical Consistency Map": it quantitatively identifies pixels that are badly performing that could be a consequence of the non-uniform application of the coating and degradation of the film, non-uniform distribution of the excitation light and occlusions due to biofouling and bubble formation. This provides a direct, real-time indicator of the sensor's spatially resolved performance and was shown to be closely aligned to the average intensity from the red channel **(Fig. 4A-B),** which on its own though could not be a good predictor in noisy environments across multiple days, as shown earlier in the first section of the Results. However, an important result was that the attention maps of both PCNN and CNN did not exhibit this alignment (**Fig. 4C-D)**. This shows the difficulty of relying solely on the neural network's output to focus on the correct regions for best performance and leaves room for improvement with other deep learning algorithms.

Trained end-to-end on raw image data using a Leave-One-(Day-)Out Cross-Validation (LOOCV) strategy, the PINN achieved a best-case-scenario test MAE of 5.4 µmol/L when the biofouling supervision term was included (PCNNB, equation 2; **Fig. 5A**) and 5.5 µmol/L without it (PCNN, equation 1; **Fig. 5B**). The negligible improvement when using the biofouling supervision is important, since it shows that external supervision using the UV-Vis data for the levels of biofouling in the water does not provide a significant benefit towards the prediction accuracy. Overall, the PCNN model resulted in a 32 **%** reduction of MAE over the best CNN baseline test MAE and a 58% reduction from the best LGBM model (13.2 µmol/L). This was achieved while being more parsimonious, learning a compact 512-feature representation of the image (versus a 48x48x3=6912 feature matrix). Additionally, the training history plot confirms the stable and effective optimization of the PCNN **(Fig. 5C)**. It illustrates the concurrent minimization of the validation loss and the physics-regularization loss for each consecutive epoch. This synergistic learning process, where improving physical consistency aids in data fitting and vice-versa, demonstrates that the model is integrating physical constraints to achieve a more generalizable solution. However, from the histograms of the prediction error, the standard deviation remained relatively high, yielding values of ~8 and 10 µmol/L for PCNNB and PCNN, respectively (**Fig. 5D-E**). Importantly, when looking at the feature importance during the Optina hyperparameter optimization procedure, the physics loss weighting factor (lambda physics) was identified as the most influential hyperparameter **(Fig. 5F)**. This underscores the critical role of physical regularization to achieve optimal model convergence and performance.

We also compared our standard approach, i.e. implementing a PINN where the network learns a single set of SV parameters for the whole frame, to a "Physics-Guided" Neural Network (PGNN) where we use pre-computed SV parameters for each pixel (following again the "Per Pixel Fitting" strategy). Specifically, in the PGNN, SV parameters were pre-fitted for each pixel via the classic calibration route and fed to the network as spatial maps. Crucially, using the PGNN resulted in a significantly higher MAE of 8.7 µmol/L. Hence, we concluded that the end-to-end PINN implementation was both superior and parsimonious. This demonstrates that allowing the network to learn effective physical parameters through a generalized SV equation per experimental day provides superior regularization and is a more robust strategy compared to fitting every pixel to pre-fitted local SV parameters.

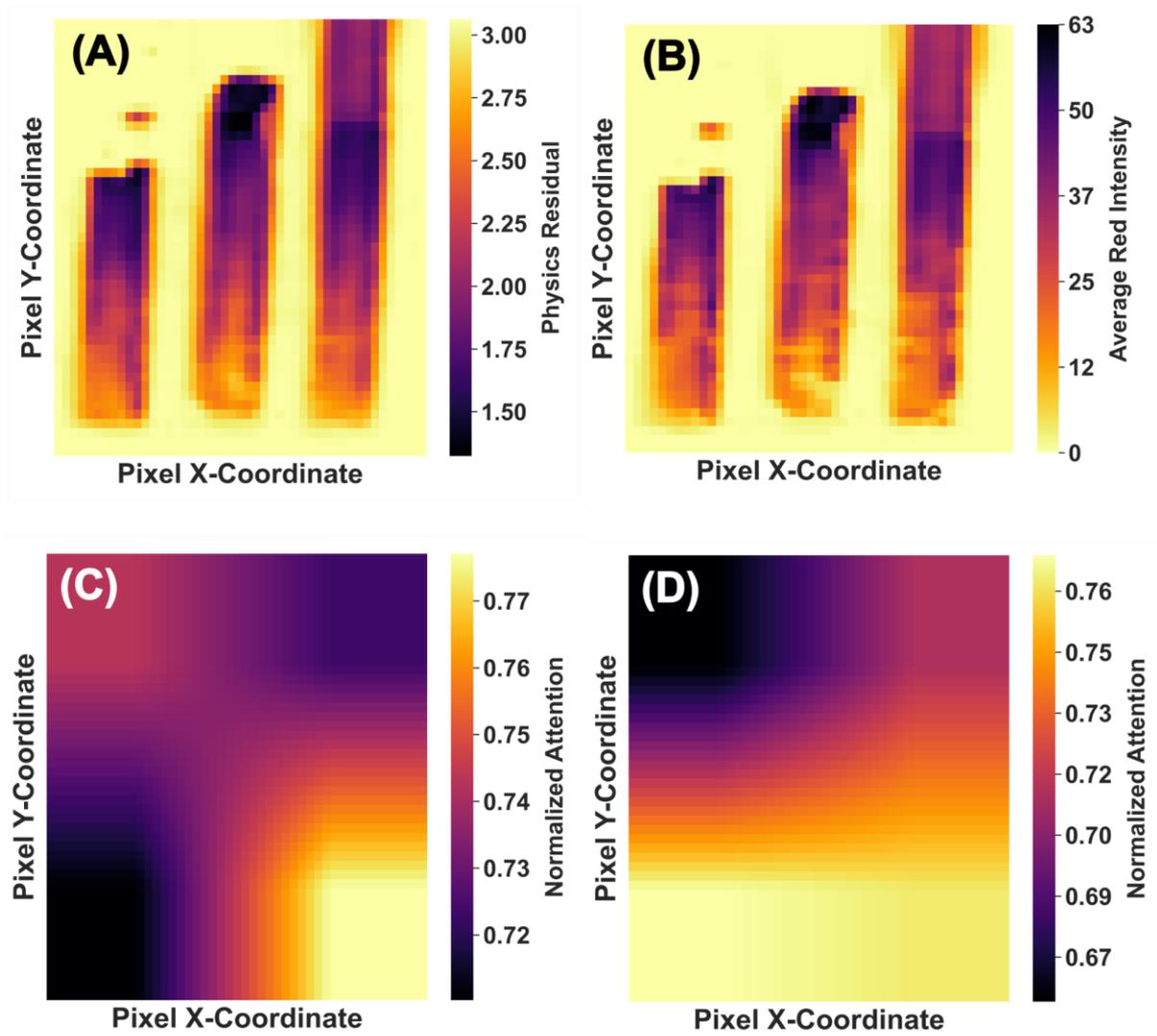

**Figure 4:** PINN Performance heatmap visualization.
Heatmaps of: **(A)** the physics Residual, **(B)** the average red intensity over the entire calibration procedure (reversed colors), **(C)** the normalised attention of the CNN and **(D)** the normalised attention of the PCNN.

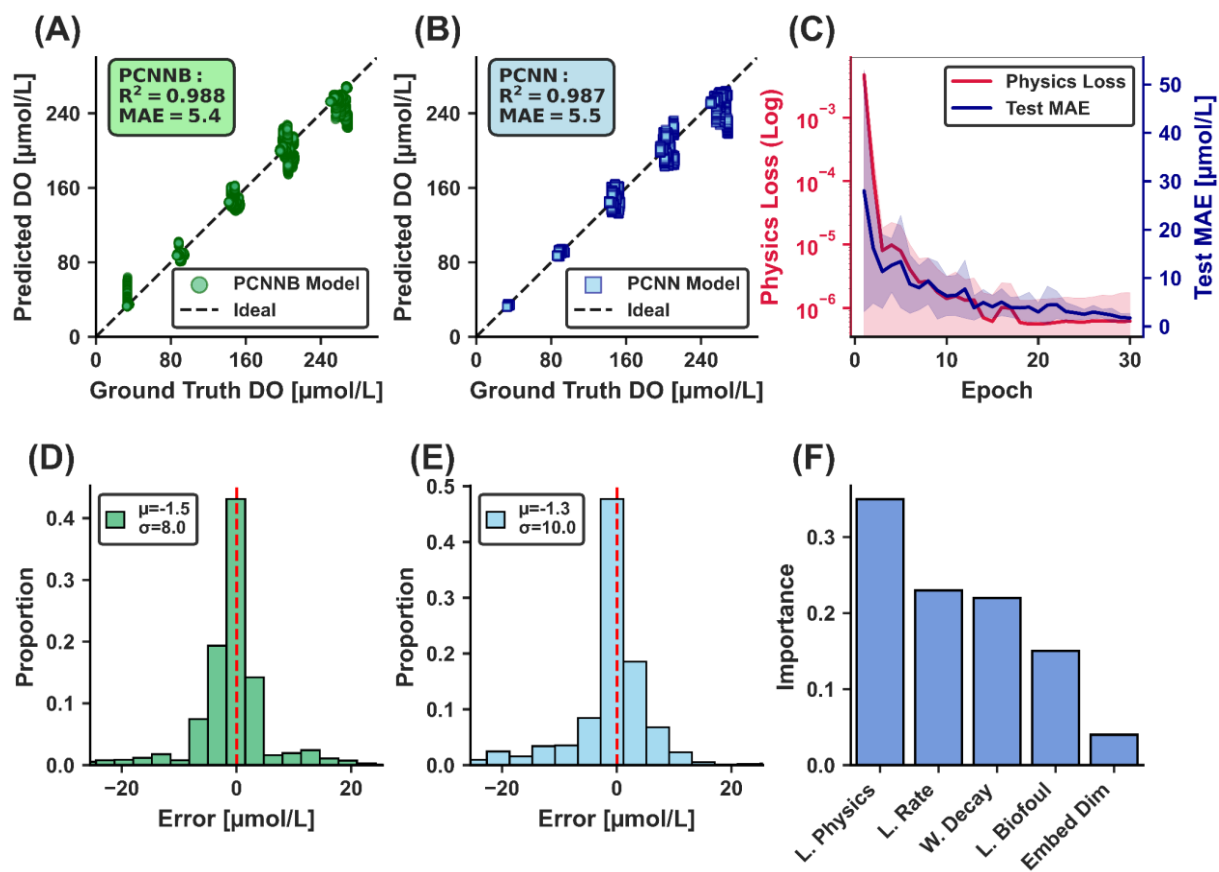

**Figure 5:** PINN Performance.
Parity plots of the CNN based PINN with **(A)** and without **(B)** the biofouling component (PCNN and PCNNB, respectively). The correlation between predicted and ground truth data is quantitatively shown with $R^2$ and MAE as the metrics of reference. **(C)** Training history plot showing the concurrent decrease of data loss and physics loss in the y axis, versus the training epoch (the lighter colours indicate the standard deviation from each experiment day). Histograms of the prediction errors for the PINN models **(D)** PCNN and **(E)** PCNNB. The y-axis shows the normalised frequency (Proportion), and the inset legend box provides the mean (μ) and standard deviation (σ) of the error. **(F)** Bar chart of the Optuna hyperparameter optimization feature importance for PCNNB.

**Visual Transformers (ViT)**
To further advance the state-of-the-art and address the potential limitations of local feature extraction in CNNs, we replaced the ResNet-18 backbone with a more powerful Vision Transformer (ViT) architecture [37,38]. Unlike CNNs, ViTs employ a self-attention mechanism allowing the model to weigh the importance of every image patch relative to all other patches simultaneously, providing a global, contextual understanding of the entire sensor surface in a single layer. This is particularly advantageous for identifying large-scale, diffuse patterns characteristic of developing biofouling that may be missed by localised feature regions. The ViT-based PINN (PViT) achieved a remarkable cross-validated best-case-scenario test MAE of 1.3 µmol/L (**Fig. 6**). This represents a further 76 % reduction in error compared to our best ResNet-based PINN and an order-of-magnitude improvement over traditional methods, establishing a new benchmark for accuracy.

Furthermore, a crucial aspect of all sensors is the ability to define the uncertainty of the sensor output. This is a fundamental bottleneck in ML-based frameworks where achieving a low average error is insufficient for a trustworthy sensor, as prediction accuracy can fluctuate significantly depending on the initialisation weights in the training of the model. This was especially true in the case of implementing ViTs. We tried to address this with a quantification of the prediction uncertainty by implementing a Deep Ensemble methodology: instead of training a single model, we trained an ensemble of 3 identical PViT models for each cross-validation fold, with the only difference being their random weight initialization around the hyperparameter optimised values (optimization was performed once more using Optuna). The low number of random initializations within the ensemble is due to the high computational resources needed for each iteration. Notably, defining the DO value as the average prediction (PViT-EA) of each member of the ensemble produced lower MAE of 1.7 µmol/L (**Fig. 6A**), while using the best choice by MAE between the ensemble members for each fold (PViT-EB) yielded an MAE of 1.9 µmol/L (**Fig. 6B**).

Importantly, for all models, and especially in the train-based MAE choice of models, some folds (experiment days that were used as the test sets) constituted large outliers, significantly affecting the MAE and the variance between the experiment days (**Fig. 6C**). This once more emphasizes the large interexperiment spatial and temporal variability of the algal biofouling. Moreover, the PViT-EA and PViT-EB models still yielded a high standard deviation of prediction errors (σ) of 2.6 and 2.8 µmol/L, respectively (**Fig. 6D, E**) compared to the MAE. This was not significantly different from the model with Optuna optimised initialisation parameters and choosing the best-case-scenario test MAE models, PViT-O, with an MAE of 1.3 µmol/L. Fundamentally, all PViTs provided lower test MAEs compared to the previous PCNN models (**Fig. 6F**), and this was true even when comparing the best-case-scenario CNN models versus the realistic models chosen from best training and validation MAEs (see next section).

Furthermore, the standard deviation of the predictions in the ensemble of each fold serves as an additional quantitative measure of the model's uncertainty. The hypothesis is that high standard deviation implies the models in the ensemble disagree, indicating a challenging or out-of-distribution sample for which the prediction should be trusted less. This was verified by a strong positive correlation between the predicted standard deviation and the actual absolute prediction error with a Pearson's Correlation Coefficient (PCC) of 91% for 3 ensemble members when training and testing on the 11 best days **(Fig. 7A),** which falls to 54% including the two experiment days with high outliers. In essence, the ensemble's uncertainty metric proved to be a reliable indicator of its performance in most experimental cases. This means the model can effectively self-identify low confidence measurements, a critical feature for real-world deployment where a sensor must be able to flag its own potentially unreliable readings. This provides a foundation for future "smart sensors". However, one issue that became apparent with the use of transformers is the higher physics loss. This indicates that for future studies different architectures and additional computational resources should be implemented to yield greater prediction performance. Despite this, there was once more a concurrent reduction in the physics loss and the validation error **(Fig. 7B)**. Importantly, we categorised the results by DO concentration, revealing that higher DO ranges correlate positively with increased prediction error (**Fig. 7C**), which was expected from the SV physics. This highlights specific DO regimes where the model faces greater difficulty. Additionally, we used a rejection curve to demonstrate that systematically discarding the predictions with the highest calculated uncertainty from the ensemble members' predictions leads to a monotonic reduction in the Test MAE, confirming that the uncertainty quantification effectively acts as a reliability filter **(Fig. 7D).**

Moreover, we evaluated the framework's scalability and data efficiency through a sequential learning analysis by steadily increasing the training days and observing the decrease in the test MAE (choosing each time the models with the lowest train MAE). Simulating an ongoing deployment, we incrementally increased the training set size from a single initial day up to 12 days. In each iteration, the model was trained on the first $N$ Days (randomly chosen), and strictly tested on the remaining unseen future days ($N$+1 to 12) **(Fig. 7E)**. Also, we replicated this with a strict chronological separation of training and testing days, while also including a validation day to guide us in choosing the optimal model. This produced a similar monotonic reduction of the MAE **(Fig. 7F)**. In both cases, we observed a rapid convergence in performance, showcasing the scalability of the proposed sensing framework based on a small number of calibration iterations prior to deployment.

Furthermore, the global nature of the ViT's self-attention mechanism proved superior for generating coherent and physically meaningful diagnostic maps. It was important showing that the two independent views, the physics residual and the attention map, tell the same consistent story. This shows that there is a strong inverse correlation between the attention weights and the physics loss; regions with high physics loss receive low attention, while regions with low physics loss receive high attention **(Fig. 7G-H)**. This synergy is the core of the self-diagnosing capability. The physics loss identifies why a region is bad (it violates the physical law), and the attention mechanism shows how the model intelligently adapts to this failure by ignoring it. The attention maps, in contrast to the CNN methodology, produced by the ViT were sharper and clearly delineated the boundaries of the films. This interpretability, combined with quantifiable uncertainty, moves beyond simple prediction and towards a truly self-diagnostic device, allowing the sensor to assess confidence in its predictions.

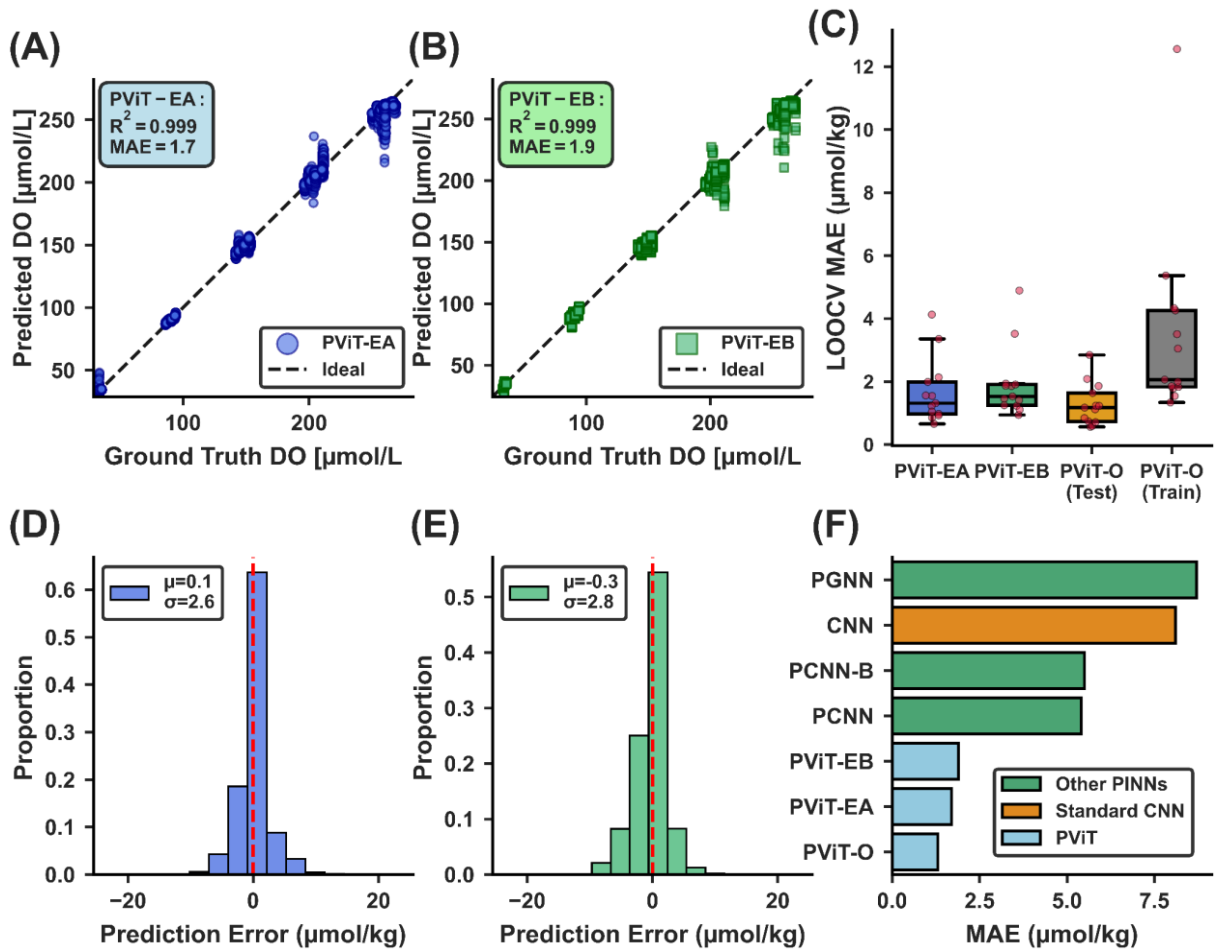

**Figure 6:** Vision Transformer Performance.
The parity plots for the **(A)** PViT-EA and **(B)** PViT-EB. **(C)** Box plots showing the distribution, median, spread (IQR), and outliers of error data for each of the 13 experimental days/folds for the 4 models: PViT-EA, PViT-EB, and PViT-O models chosen for best test and train MAE. The histogram of the prediction errors for models **(D)** PViT-EA and **(E)** PViT-EB. The y-axis shows the normalised frequency (Proportion), and the inset legend box provides the mean (μ) and standard deviation (σ) of the error. **(E)** A summary bar chart showing the MAE for the best performing models: (from left to right): PViT-EB, PViT-O (test), PViT-EA, PCNN-B, PCNN, CNN, PGNN.

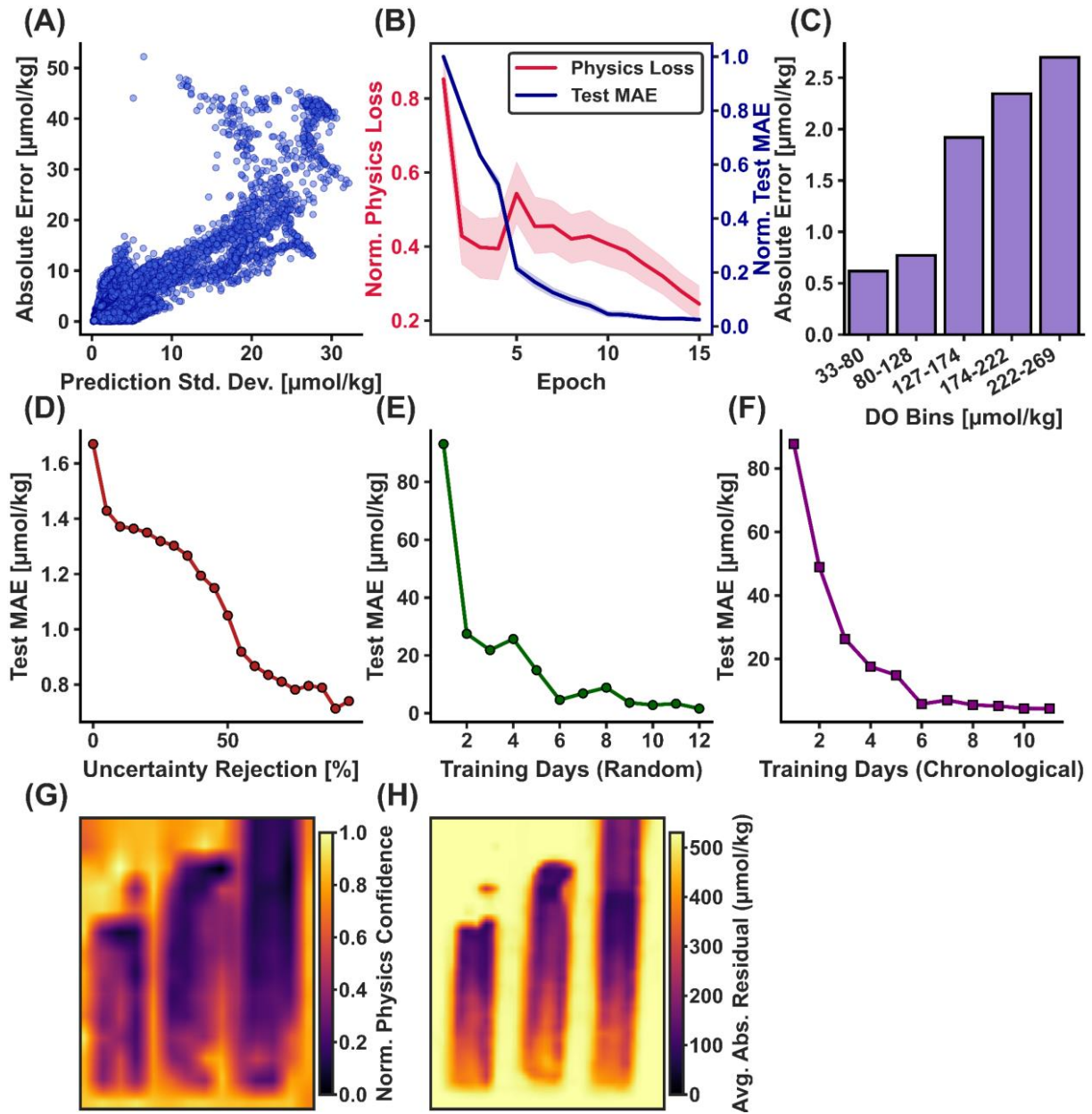

**Figure 7:** ViT uncertainty of the prediction error.
(**A**) A scatter plot of the predicted uncertainty from the model PViT-EA (standard deviation of predictions) versus the absolute prediction error for every pixel. (**B**) Training history plot showing the concurrent decrease of validation loss and physics loss in the y axis, versus the training epoch (the lighter colours indicate the standard deviation from each experiment day). (**C**) Absolute error of the prediction shown as bars for specific oxygen bins averaged over all the ensemble members. (**D**) Improvement of the test MAE based on the rejection of high uncertainty measurements from the ensemble members' predictions (from 0 to 95% rejection). (**E**) The lowest test MAE achieved plotted versus the number of experiment days used for training (chosen randomly for each iteration), indicating the scalability of the proposed framework. (**F**) The lowest test MAE achieved plotted versus the number of experiment days used for training, implementing a strict chronological order and using one day as validation each time. A side-by-side comparison of the physics confidence mask (**G**) and the averaged Physics Residual Maps generated by the PViT (**H**).

**Robustness to Temporal Extrapolation and Independent Configurations**

To assess the viability of a sensing framework for realistic "deploy-and-forget" applications, a strict temporal forecasting methodology was further investigated. In essence, for a field application, the sensor must be able to infer the DO concentration based on a model that has been trained on chronologically previous data using the ML framework proposed. To show this, it is important that DO levels are inferred for chronologically future time-points based solely on historical calibration data. Up to now, the majority of calculations (exception is **Fig. 7F**) presented had separate training and testing samples where one test is left out for testing (LOOCV), and the MAE reported was the lowest test MAE achieved after training with a number of epochs (25-35). Hence, it is crucial to also present the results for a strict chronological split (training/test) where we use the first 11 days for training, validate on day 12 and test on the unseen final day (day 13). In this scenario, the PViT, once more, successfully extrapolated the DO values, maintaining a low MAE of 2.5 µmol/L and an $R^2$ of 99.5%. Moreover, we developed and applied the model using LOOCV for all days, this time with 11 days for training and 1 for validation, and the average test MAE across all days was 2.0 µmol/L (**Fig. 8 A, B**). This, in combination with the fact that the lowest validation and training MAE correspond to the same model (given the best choice with an ensemble methodology) for every experimental day, indicates that this framework allows for the optimization of a model that can successfully infer DO concentration in biofouled environments. Additionally, by choosing a model without the use of a validation set and without the ensemble technique, the MAE was higher at 3.5 µmol/L, indicating that the combination of the ensemble and validation set is beneficial in the more realistic scenario.

Nonetheless, it is pivotal to highlight the generalizability of this new sensing paradigm. Thus, we validated the framework's transferability through a completely independent experiment involving a single newly fabricated PtOEP film, a different UV-LED configuration, a continuous DO concentration range, a temporally dynamic DO, and unfertilised algae over a 4-day period. This further proves the adaptability and resilience of the proposed framework to different setups and conditions by testing it on a unique set of film heterogeneities, DO concentration distributions and a different biological growth rate. Once more, we use a strict chronologic-based splitting of the data into training and testing, thus we trained the model on the first 3 days and tested on the 4[th]. The validation set in this case was 5% of the total sampling pool, randomly selected, not part of the training set and chronologically before the test samples. We used the same DO and temperature sensors.

In this case, both the PCNN and PViT models demonstrated remarkable resilience despite the small training dataset. The PViT and PCNN models achieved an MAE of 4.4 µmol/L (**Fig. 8C**) and 4.8 µmol/L (**Fig. 8D**), respectively, significantly outperforming the Classical approach (21.4 µmol/L) (**Fig. 8E**). Once more, the higher DO levels incurred the highest errors, just as expected due to the SV physics and previous results (e.g. **Fig. 7C**). It should be noted that the slightly higher overall MAE compared to the previous experimental configuration was expected due to a) smaller training datasets, b) the use of one film, c) larger range of DO concentrations, and d) higher temporal fluctuations. Despite the increased MAE, these results confirm that the proposed architecture is generalizable for DO concentration measurements in water via quenching of phosphorescence with high accuracy.

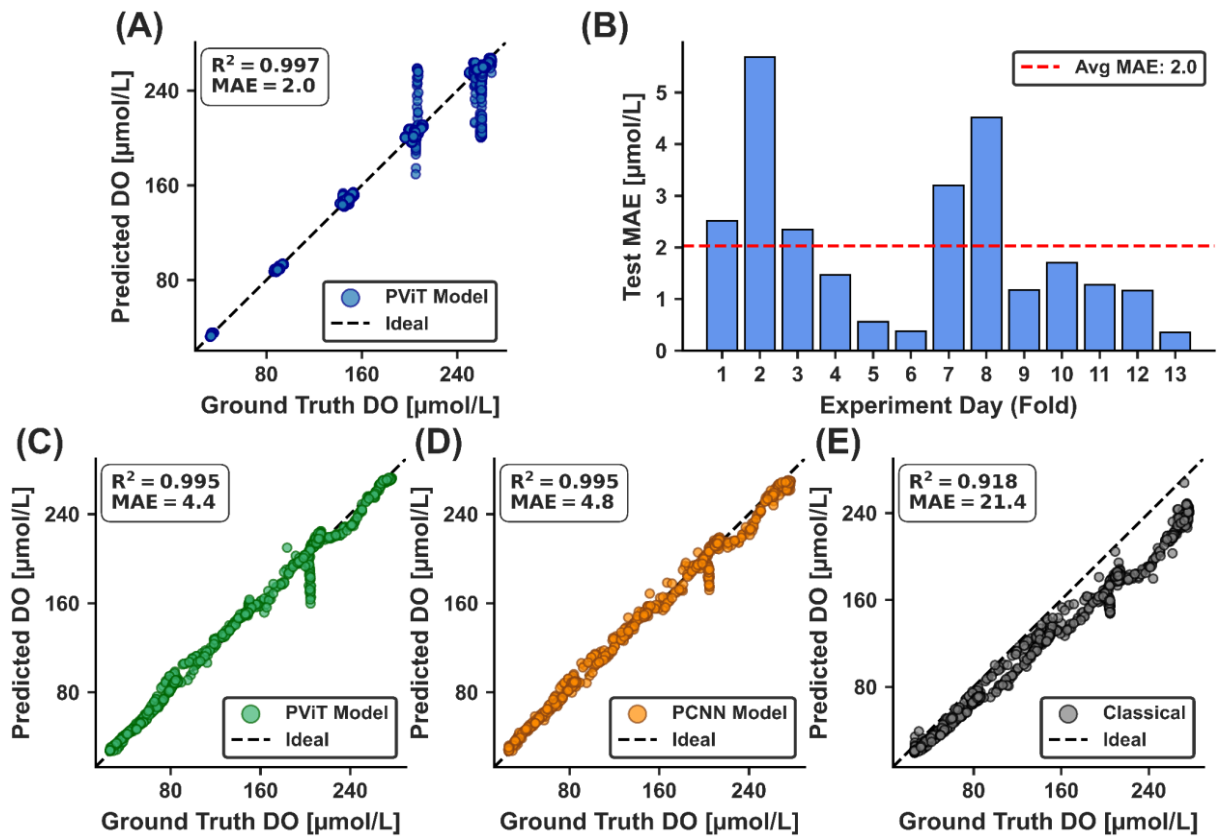

**Figure 8:** Robustness to temporal extrapolation and generalizability to independent experimental setups:

**(A)** Parity plot illustrating the PViT model's performance in a strict temporal forecasting scenario, where the model was trained on historical data (Days 1–11), validated on Day 12 and tested on the unseen final day (Day 13). **(B)** Bar plot displaying the Test MAE for the PViT model obtained for each of the 13 experimental days using a validation set and Leave-One-Out Cross-Validation (LOOCV) strategy, demonstrating consistent performance across the dataset. The bottom row evaluates the framework's transferability in a completely independent "blind" experiment involving a new sensor film and setup. Parity plots compare the prediction accuracy on the unseen test day (Day 4) for the PViT model **(C)**, the PCNN model **(D)**, and the best case of the Classical calibration ("Best Pixels") approach **(E)**.

## Discussion

In this work, we demonstrate a significant advancement of in-situ optical sensing of DO. We show that the vulnerability of luminescence-based sensors in real-world environments lies not inherently in the materials, but in the calibration techniques and sensing methodology. The transition to imaging modalities exposes the fundamental limitation of assuming spatial homogeneity in both non-imaging approaches (such as in industry practice) and in the average-based calibration strategies. However, simply expanding the sensor's spatial resolution but still relying on traditional averaging or static "best-pixel" selection failed catastrophically to predict the DO concentration (MAE $\geq$ 18.5 µmol/L). The same was true for position agnostic feature-based ML approaches with a significantly larger input matrix (MAE $\geq$ 34.6µmol/L), and even spatial feature LGBM models (MAE $\geq$ 13.2 µmol/L).

On the other hand, by using architectures that consider spatial correlations between pixels and by embedding the governing SV equation directly into the deep learning process, we arrived at a framework that is both accurate and interpretable. Specifically, a Vision Transformer (ViT) with a physics loss component achieved a state-of-the-art MAE of ~2 µmol/L (when training and testing datasets were chosen in chronological order) under accelerated fouling over a 14-day period. Additionally, using a Deep Ensemble methodology combined with the ability to generate spatial heatmaps of physics loss and attention provide a dual mechanism for sensor self-diagnostic capabilities. This enables future systems to flag not just where they are rapidly degrading (via physics residuals) but also how much to trust their current output (via uncertainty). We were also able to show the generalizability of the proposed framework by applying it on a similar setup with a different configuration and different conditions. Together, these advances yield a sensing platform that remains robust even under fouling conditions.

By combining low-cost, self-assembled films and accessible hardware with a powerful computational framework, we democratize access to high-performance environmental monitoring. Crucially, the proposed platform achieves a ~50-120x improvement in cost-efficiency compared to current industrial devices and is readily scalable; envisioned for a 'deploy-and-retrieve' model where the entire sensor is housed in a self-contained package and deployed for a monitoring period and then collected for data offloading and refurbishment. This strategy aligns with the low-cost nature of the hardware, simplifies field logistics, eliminates the need for complex in-situ data transmission and allows for IOT (Internet-Of-Things) devices. Thus, large-scale network deployments on autonomous underwater vehicles become feasible, which in turn enable accurate, long-term ocean monitoring that is essential for forecasting climate tipping points. Future work should focus on oceanic deployment for longer periods to assess long-term stability and on-board self-calibration techniques using IOT devices.

## Methods

**Film Fabrication**

The oxygen sensing film creation has been described in depth in a previous publication [33] and relies on self-assembly fabrication using the Breath Figure technique with Polystyrene as the encapsulation polymer and PtOEP as the oxygen sensitive dye.

**Experimental Setup**

We used a Raspberry Pi, an HD Pi Camera, a UV LED (395nm – Wurth Elektronik), and Pyroscience DO sensor (Pico-O2-SUB) as the ground truth sensor measurements. The 14-day algae growth experiment was performed by purchasing one type of algae (Chlorella) along with the appropriate fertilizers. The industry sensor was cleaned every 2 days with DI water, inserted into the same position and was at the same height in the water enclosure as the placement of the films. The biofouling was not only related to the darkening of the pixels as seen from the camera because most of the occlusions occurred on the side of the film that was not facing the camera or the UV LED and the film is opaque (**Fig. 1**) (see also the biofouling quantification section below). The positioning of the film compared to the UV LED and camera was chosen to mirror the setup of industrial sensors. The temperature was monitored using the DB9 temperature sensor connected to the Raspberry Pi. From the 14 day-period, 13 datasets from 13 separate days were used to train and test the model. The proposed optoelectronic configuration achieved a ~50-120x cost improvement factor compared to current industrial solutions.

All neural network architectures, including the CNN, PINN, and ViT models, were implemented using the PyTorch deep learning framework. To handle the high computational demands of the Vision Transformer and the physics-informed loss calculations, model training and inference were hardware-accelerated using an NVIDIA GeForce RTX 5070 GPU with 12 GB of dedicated VRAM. In some instances, procedures were accelerated using NVIDIA A100 GPUs on the University College London (UCL) computing cluster Myriad. Training was further accelerated using Automatic Mixed Precision (AMP) to optimise memory consumption.

**Experimental Methodology**

To establish a robust performance baseline and demonstrate the limitations of existing approaches, we first evaluated a spectrum of conventional and machine learning calibration techniques using only the red-channel pixel intensities, which most closely correspond to the emission spectrum of PtOEP. Then, CNN and ViT-based models were implemented and all 3 RGB channels were included. Each video frame was downsampled to 2 fps and 48x48 pixels from 30 fps and 1920x1080, respectively, due to the computational resources available and the processing time of the frames. For context, optimizing a single ViT model using LOOCV required approximately 7 minutes per training epoch (totalling roughly 3 hours for a complete 25-epoch cycle) with the hardware available (NVIDIA GeForce RTX 5070 12 GB VRAM). This corresponds to training with 12 days of historical calibration data (~24,000 frames) and testing on 1 day (~2000 frames).

The ground truth data was obtained in parallel from a DO sensor (O2 Pico-Sub Pyroscience) and the calibration was performed in 5 stable DO concentrations achieved using mass flow controllers set at 0, 5, 10, 15, and 20 V/V %. During the 14-days of algae growth, we performed a calibration each day and used the frames from 13 videos (each from a separate day) for training and testing the models. To reduce the random systematic error from the industry sensor we resampled the ground truth data using their average every 10s (step averaging), which is expected to be true for a constant flow from the mass flow controllers (with an accuracy of 0.1% V/V). This was a critical part of the conditioning of the input ground truth signal to avoid optimizing the ML model on the Pyroscience sensor noise.

All methods were assessed using a stringent LOOCV protocol [39], where each calibration from a separate day was held out as an unseen test set and corresponded to a different set of weights for the final ML model. This constitutes the closest scenario to a real-world application where we would employ a new calibration methodology to predict an entire unseen dataset (corresponding in this case to a single day/calibration video) and thus was considered the most relevant way of organising the testing and training of the model. Fundamentally, the nature of algae growth on the surface of the sensor is not homogeneous in time or space and thus the average of all possible combinations of days for training and testing is a more accurate depiction of the implementation of this sensing framework. Nonetheless, a purely chronological analysis has also been included. Additionally, we highlighted the scalability of the proposed framework by splitting the dataset into training and testing with 10 different ratios (keeping the total the same, 13), using randomly selected experimental data sets of each day, to measure the effect of training size versus testing on the final test MAE, as a metric of scalability (**Fig. 7E**).

**Realistic Temporal Validation and Independent Experimental Verification**
The models were initially compared in the idealised scenario by choosing to report the best test MAE achieved over 35 training epochs, with the goal of reporting the best possible results to compare with the best ViT model. While the LOOCV provides a robust baseline, the metric of best test MAE performance from a set number of epochs does not directly translate in a realistic deployment scenario. Thus, we further validated the framework's ability to generalize temporally under strict "deploy-and-forget" conditions, where the model must predict future sensor performance based solely on chronologically past data.

To achieve this, we first re-evaluated the primary dataset using a strict chronological split: the first 11 experimental days were used for training, the 12th day served as a validation set for hyperparameter tuning and early stopping, and the 13th (final) day was reserved as a completely unseen test set. Notably, unlike the LOOCV approach where the best model might be selected based on test performance, here the model selection was determined exclusively by the training and validation set metrics. This ensures no "data leakage" from the future occurs, simulating a realistic scenario where the sensor must extrapolate the biofouling trend without access to future calibration points. Additionally, this was replicated with all possible number of training days to measure how fast the model would converge to lower test MAEs based on the available training days, again using a validation set (of a single experiment day) directly after the training dataset.

Crucially, to rigorously test the transferability of the framework to a completely new physical setup, we also conducted a second, independent 4-day experiment using a newly fabricated sensing film, a different setup configuration, and a fresh batch of unfertilised algae. This "blind" experiment was designed to assess the framework's robustness against variations in film fabrication and distinct biofouling growth rates. The dataset was split chronologically, using the first 3 days for training and the 4th day for testing. For model selection, we reserved a 5% uniform sample from the training days as a validation set. We applied the Classical global averaging, PCNN, and Vision Transformer (PViT) models to this independent dataset. This approach confirms that our results are not artifacts of overfitting to a specific experimental realization but represent a generalised capability of the physics-informed framework in tandem with the transformer architecture to adapt to novel environments and unseen sensor films.

**Biofouling Quantification**

We quantified relative biomass using crystal violet (CV) staining on seven coated acrylic "sister" samples collected every two days. Each sample was (i) rinsed in deionised (DI) water to remove loosely attached algae, (ii) fixed in glutaraldehyde to preserve the formed biofilm, (iii) stained with CV for 2 min, and (iv) rinsed again with DI water. The bound dye was then eluted by submerging the sample in a vial of alcohol for 1 h. The resulting dye/alcohol solutions were analysed by UV–Vis spectroscopy **(Fig. 9).** All solution volumes (rinses, fixative, stain, and solvent) were kept constant across samples to ensure reproducibility.

The Crystal Violet staining in conjunction with UV-Vis measurements were a clear indication that organic biomass adhered onto the film and increased in size throughout the duration of the experiment (**Fig.9D**). Specifically, the Crystal Violet staining was used to quantify the biofilm formation, as a thin membrane adhered on top of the phosphorescent film, and did not quantify the optically-occluding macro-aggregates (large pieces of visibly green algae), which were rinsed off the surface of the samples with DI water.

This was different to the visualisation capabilities of the camera. In detail, due to: i) the configuration of the setup where the phosphorescent film is placed on the other side of the camera and is observed through a UV-transparent acrylic, ii) the opaque nature of the phosphorescent film, and iii) the transparency of the biofilm; the camera could not visualize directly the biofouling on the sensor surface, but only the occlusions on the uncoated acrylic. Moreover, the occlusions in the form of macro-aggregates and bubbles differed depending on the day of experiment and the specific frame within the calibration video (as confirmed from visual inspections). Thus, there was no gradual and steady increase observed in the regions that the biofouling and confidence masks "avoided" during the 14-day experiment like in the case of the Crystal Violet measurements. In essence, the masks were able to capture the average biofouling levels of the sensor films for each day of the experiment, by visualizing the performance of the pixels from the combination of visual occlusions and phosphorescence emitted.

Thus, for future studies, we suggest that two separate cases are investigated independently: i) the biofouling occurs on the film and is directly visualised by the camera, as the film and the camera are placed directly facing each other, and ii) the film will be on the other side of the transparent acrylic compared to the camera (as it was during this experiment), but measures are taken so that no occlusions occur between the camera and the film - with only the transparent acrylic existing in the space between them. In the latter, we essentially eliminate the ability of the camera to "visualize" the occlusions on the surface of the film prior to any DO calibration. In addition, spectroscopic techniques should be investigated using different wavelengths of LED to potentially identify the biofouling directly, which could potentially increase the prediction accuracy of the model.

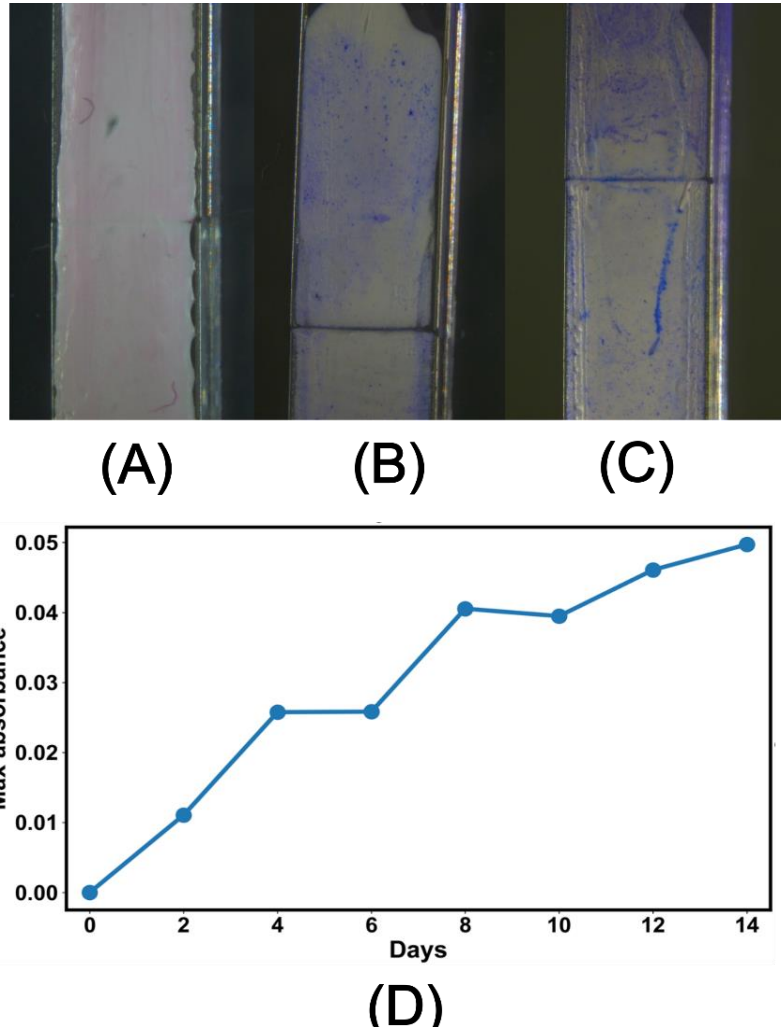

**Figure 9:** Biofouling quantification.
Images of the film applied on top of the transparent UV acrylic **(A)** without CV staining, **(B)** with staining from the second day of fouling and **(C)** the last day of algae growth. **(D)** A plot from the UV-Vis measurements of each sample (per 2 days) where the max absorbance is plotted against the day of the experiment.

**Physics derived equations**

### 1. Stern–Volmer Relationship and Sensitivity ($K_{SV}$)

The sensitivity of the sensor is defined by the Stern–Volmer constant, Ksv, derived from the linear form of the SV equation. It represents the slope of the relationship between the quenched intensity ratio and the oxygen concentration:

$$I_0 / I = 1 + K_{SV} \cdot [O_2] \quad (4a)$$

- $I_0$: phosphorescence intensity in the absence of oxygen ($[O_2] = 0$)
- $I$: phosphorescence intensity at a given oxygen concentration $[O_2]$
- $K_{SV}$: Stern–Volmer quenching constant (µmol$^{-1}$·L), a measure of sensitivity

For heterogeneous environments, a non-linear two-site model is often used to account for multiple quenching domains with different sensitivities:

$$\frac{I_0}{I} = f_1 \left(1 + K_{SV,1} \cdot [O_2]\right)^{-1} + f_2\left(1 + K_{SV,2} \cdot [O_2]\right)^{-1} \ (4b)$$

- $f_1, f_2$: fractional contributions of each site ($f_1 + f_2 = 1$)
- $K_{sv,1}$, $K_{sv,2}$: Stern–Volmer constants for each quenching site

if you replace $f_2$ and $f_2$ (the fractional components) with a single parameter $a$ representing the fractional contribution of the first component, and $(1 - a)$ for the second, the Stern–Volmer relationship becomes:

$$\frac{I_0}{I} = a(1 + K_{SV,1}[O_2])^{-1} + (1 - a)\left(1 + K_{SV,2}[O_2]\right)^{-1} \ (4c)$$

where:

- $a \in [0,1]$ is the fractional contribution of the first component, with $(1 - a)$ representing the second.

**2. Dynamic Range (DR)**

The dynamic range is the total change in signal intensity across the measured concentration range, indicating the operational span of the sensor.

$$Dynamic\ Range = I_{max} - I_{min} \ (5)$$

- $I_{max}$: mean intensity at the lowest oxygen concentration (typically approaching $I_0$)
- $I_{min}$: mean intensity at the highest oxygen concentration measured

**3. Maximum Intensity ($I_0$)**

The maximum, or unquenched, intensity is a fundamental parameter representing the sensor's baseline signal in an oxygen-free environment.

$$I_0 = I \ \ as \ \ [O_2] \to 0 \ (6)$$

- In practice, $I_0$ is determined experimentally at zero oxygen concentration or estimated as a fitting parameter from the SV equation.

**4. LOD (Limit of Detection)**

The Limit of Detection is the lowest oxygen concentration that can be reliably distinguished from a zero-concentration sample. Following IUPAC, it is calculated as:

$$LOD = \frac{(3 \cdot \sigma)}{m} \ (7)$$

- $\sigma$: standard deviation of repeated intensity measurements at zero (or near-zero) oxygen concentration

- m: sensitivity (slope) of the calibration curve ($m = K_{SV}$ for the SV model)

**5. R² (Coefficient of Determination)**

The coefficient of determination, $R^2$, quantifies the goodness-of-fit of the linear SV model to the experimental data.

$$R^2 = 1 - \frac{\Sigma(y_i - \hat{y}_i)^2}{\Sigma(y_i - \bar{y})^2} \quad (8)$$

- $y_i = I_0/I - 1$ : observed value
- $\hat{y}_i = K_{SV} \cdot [O_2]$ : predicted value from the fit
- $\bar{y}$: mean of observed y values
- $n$: number of data points

**CNN training**

For the CNN architecture patterns (see Supplementary Information section S4 for full details). we used a ResNet-18 [40] backbone and a Convolutional Block Attention Module (CBAM) [41] block to highlight informative regions. The architecture leverages the ResNet-18 backbone, pre-trained on ImageNet, to extract rich hierarchical features from the raw 48x48x3 pixel sensor images. We also integrated a CBAM to "guide" the model to learn and spatially focus on the most informative regions of the sensor film, dynamically adapting to heterogeneous fouling. This network processes the full spatial information of the sensor frames, alongside ambient temperature, through a final regression head to predict the dissolved oxygen concentration. For the training we used AdamW optimizer [42] with MSE loss, automatic mixed precision (AMP), gradient accumulation and clipping, optional cosine LR scheduling, early stopping for faster results, and data augmentation (random horizontal and vertical flips). Hyperparameters (learning rate, weight decay, backbone, physics loss-when applied) are tuned with Optuna for a subset of the entire data set (30%). Evaluation is once more rigorous using LOOCV (entire experiment days held out). This protocol rigorously tests the model's ability to generalize to completely unseen temporal conditions and fouling states, preventing any data leakage between training and testing.

**Vision Transformer Training and Uncertainty Quantification**

To capture global spatial dependencies beyond the local receptive fields of CNNs, we replaced the ResNet-18 backbone with a Vision Transformer (ViT, vit_small_patch16_224) [38], pre-trained on ImageNet (see Supplementary Information section S5 for full details). The ViT divides each input frame into a grid of patches and uses self-attention to weigh every patch relative to all others simultaneously, providing a holistic representation of the sensor surface, which is particularly advantageous for modelling diffuse biofouling patterns. The global CLS token is concatenated with the temperature reading and passed to a regression head for DO prediction, while the spatial patch tokens are decoded into the biofouling mask and physics confidence map. The architecture comprises four specialised heads whose functions are summarised in Table S1 in the Supplementary Information document. The network was trained with the dual-objective loss (equation 1), using AdamW with hyperparameters tuned via Optuna; images were upsampled from 48×48 to 224×224 via bilinear interpolation to meet the ViT input requirements. To quantify predictive uncertainty, we implemented a deep ensemble methodology [43]: for each cross-validation fold, an ensemble of identical PViT models with different random weight initialisations was trained, with the ensemble mean serving as the final prediction and the standard deviation as a direct measure of model confidence.

The proposed sensing paradigm is well suited for a distributed Cloud-Edge IoT architecture: the computationally intensive Deep Ensemble ViT training is executed on centralized, high-performance hardware and once the robust, physics-informed weights are finalized, the static model is compiled and pushed to a microcontroller (e.g. Raspberry Pi) at the edge. The edge device is only responsible for Inference (a forward pass), which is highly efficient, allowing the edge device to perform real-time, self-diagnostic DO measurements and spatial uncertainty quantification locally, significantly reducing the bandwidth needed to transmit data.

## Data availability

The raw image data, processed datasets, and ground truth measurements required to reproduce the findings presented in this study are available from the corresponding author upon request and available to download from OSF.

## Code availability

The code with the models used in this work are accessible on GitHub.